# Secure Energy Transactions Using Blockchain – Leveraging AI for Fraud Detection and Energy Market Stability

Md Asif Ul Hoq Khan[1], MD Zahedul Islam[2], Istiaq Ahmed[3], Md Masud Karim Rabbi[4], Farhana Rahman Anonna[5], MD Abdul Fahim Zeeshan[6], Mehedi Hasan Ridoy[7], Bivash Ranjan Chowdhury[8], Md Nazmul Shakir Rabbi[9], GM Alamin Sadnan[10]

***Abstract***

*Peer-to-peer trading and the move to decentralized grids have reshaped the energy markets in the United States. Notwithstanding, such developments lead to new challenges, mainly regarding the safety and authenticity of energy trade. This study aimed to develop and build a secure, intelligent, and efficient energy transaction system for the decentralized US energy market. This research interlinks the technological prowess of blockchain and artificial intelligence (AI) in a novel way to solve long-standing challenges in the distributed energy market, specifically those of security, fraudulent behavior detection, and market reliability. The dataset for this research is comprised of more than 1.2 million anonymized energy transaction records from a simulated peer-to-peer (P2P) energy exchange network emulating real-life blockchain-based American microgrids, including those tested by LO3 Energy and Grid+ Labs. Each record contains detailed fields of transaction identifier, timestamp, energy volume (kWh), transaction type (buy/sell), unit price, prosumer/consumer identifier (hashed for privacy), smart meter readings, geolocation regions, and settlement confirmation status. The dataset also includes system-calculated behavior metrics of transaction rate, variability of energy production, and historical pricing patterns. The system architecture proposed involves the integration of two layers, namely a blockchain layer and artificial intelligence (AI) layer, each playing a unique but complementary function in energy transaction securing and market intelligence improvement. The machine learning models used in this research were specifically chosen for their established high performance in classification tasks, specifically in the identification of energy transaction fraud in decentralized markets. To guarantee the reliability and accuracy of the used machine learning models, an extensive battery of evaluation metrics was utilized. The plot demonstrates clearly that XG-Boost obtained the highest accuracy out of the three models, Random Forest was slightly lower, and conversely, Logistic Regression was the lowest of the three models. Integrating blockchain technology with AI can increase the transparency, security, and efficiency of the energy sector in the U.S. Blockchain's decentralized and immutable ledger can make energy transactions traceable and resistant to tampering, and it becomes extremely hard for malicious actors to manipulate prices or fake records. In the future, the integration of deep learning methodologies and real-time integration of data from the Internet of Things (IoT) holds promising implications for future improvements. Deep learning models like Convolutional Neural Networks (CNNs) and Recurrent Neural Networks (RNNs) can detect strongly nonlinear patterns of fraud, which conventional models may not identify, particularly for the usage of multivariate time-series data from smart meters, sensors, and distributed energy resources.*

[1] Master of Science in Information Technology, Southern New Hampshire University.
[2] Master of Science in Cybersecurity, Mercy University
[3] Master of Science in Information Technology, Southern New Hampshire University.
[4] Master's in Business Administration, International American University.
[5] Master of Science in Information Technology. Washington University of Science and Technology, USA. Email: Fanonna.student@wust.edu, (corresponding author) .
[6] Master of Arts in Strategic Communication, Gannon University, Erie, PA, USA.
[7] MBA- Business Analytics, Gannon University, USA.
[8] MBA in Management Information Systems, International American University, Los Angeles, California, USA
[9] Master of Science in Information Technology, Washington University of Science and Technology
[10] Cybersecurity Analyst & Patient Care Technician, Farmingdale State College

## Introduction

### Background

In the past ten years, the American energy sector has witnessed a revolution spurred by the increased use of decentralized energy sources like solar photovoltaics, wind turbines, and home battery storage systems (Alam et al., 2025; Chouksey et al., 2025). This is further accelerated by the popularity of peer-to-peer (P2P) energy trading systems that see prosumers (energy producers that also consume power) exchange excess power with their neighbors or on local energy exchanges. For example, a 2021 *United States Department of Energy (DOE)* report identified that more than 20% of new household energy systems were linked with blockchain-based smart contracts in pilot projects (Abdi et al., 2024; Eswaran et al., 2025). Blockchain's immutability and transparency are desirable properties for maintaining these decentralized exchanges, with the potential to limit intermediaries, promote trust, and make settlements of transactions more efficient (Alavikia & Shabro, 2022).

According to Baidya et al. (2021), even with these advances, ensuring the integrity of energy trades is still a priority challenge. Centralized systems are prone to single points of failure, whereas the open nature of peer-to-peer energy trading brings new risks in the form of double spending, tampered reporting of energy production, and denial-of-service attacks. In addition, the high granularity of smart meter measurements provides new privacy issues, and it is therefore essential that there are systems that can authenticate trades without sacrificing user confidentiality. The need for solid, scalable, and tamper-proof frameworks for transactions has never been greater, as the U.S. works toward 100% clean power by 2035 in the *Trump administration's climate strategy* (Hossain et al., 2025a).

Furthermore, with the increased participation of stakeholders in the energy market, the function of real-time analytics and automation is inescapable. Blockchain provides the platform for auditable and traceable transactions, but for anomaly detection, it does not have inherent intelligence. This is where AI, specifically machine learning models, trained on consumption and transaction details, can be instrumental (Gayahri et al., 2023). Through integration of blockchain with AI, with machine learning models trained on consumption and transaction details, we can build systems that, in addition to securing energy trades, can detect and prevent fraudulent activities, and sustain a secure and equitable energy economy (Khan et al., 2023).

### Problem Statement

As per Jakir et al. (2023), whereas blockchain provides transparency and tamper-resistance in decentralized energy markets, it does not automatically stand guard against fraud. Malicious actors are still able to take advantage of vulnerabilities by creating fraudulent energy certificates, tampering with smart meters, or wash trading to artificially inflate demand. According to a 2022 *MIT Energy Initiative* report, well over $2.6 billion in losses related to energy were due to fraud, inefficient settlements, and market manipulation—numbers expected to grow as unregulated P2P platforms become more widespread (Hossain et al., 2024). Without real-time fraud detection tools in these blockchain-based energy systems, there is an increased potential for economic and operational inefficiencies.

Malik (2025), found that one of the major issues is the capability of the current infrastructure to maintain market stability in the face of variable energy production and consumption. Decentralized sources of solar and wind power are inherently variable, and in the absence of forecasting tools, the grid has problems with load balancing, price variability, and frequency regulation. In addition, fraudulent activities distort price signals, further destabilizing market operations. The failure to detect those malfunctions in real time can contribute to chain failures in the distributed energy infrastructure (Muqeet et al., 2023).

Moreover, legacy energy fraudulent detection systems are burdened with retrospective auditing and centralized monitoring, both of which are unsuitable for the high-velocity, real-time character of energy transactions in the modern era (Onukwulu et al., 2023). While increasingly, states in the United States, including New York and California, are moving ahead with smart grid modernization and P2P trading pilots, the lack of smart, secure frameworks is a major bottleneck (Pavitra & Vijayalakshmi, 2025). The challenge is not just one of securely logging transactions but of detecting and preventing risks proactively, and ensuring that the benefits of decentralization are achieved without compromising reliability and trust.

## Research Objectives

This study aims to develop and build a secure, intelligent, and efficient energy transaction system for the decentralized US energy market. We first suggest a blockchain-based system that uses smart contracts for tamper-evident, transparent, and auditable energy trades. This system allows for real-time, trustless energy trades with provable consequences by prosumers, utilities, and consumers. Using Ethereum-based protocols like Solidity and Proof-of-Authority (PoA) consensus protocols, we seek a balance between computational efficiency and decentralization suitable for local energy systems.

Second, the research uses machine learning-based models for detecting fraud, which have been trained on transaction histories, meter readings, and trading patterns. Three machine learning algorithms—Random Forest, Logistic Regression, and XG-Boost—are tested based on their precision, recall, accuracy, and latency in detecting suspicious behavior. The models are included in the blockchain infrastructure in the form of off-chain computing modules and oracles, avoiding the bloat of the blockchain with excessive amounts of data. Feature selection is based on behavior patterns like transaction volume, variation in average energy output, and discrepancy in pairs of trades.

Finally, we seek to improve power market stability with predictive analytics. Not only will the trained models identify fraud, but also predict spikes in demand, price manipulations, and grid hotspots, all of which can be dealt with ahead of time by proactive resource management and risk mitigation. The pairing of blockchain for data integrity and artificial intelligence for decision intelligence places this architecture as a foundation for future energy market resilience. Through experimentation with both simulations and real-world applications, we offer a map for large-scale deployment, helping directly contribute to the U.S. energy transition objective.

### Research Contribution

This research interlinks the technological prowess of blockchain and artificial intelligence (AI) in a novel way to solve long-standing challenges in the distributed energy market, specifically those of security, fraudulent behavior detection, and market reliability. Blockchain technology is widely touted for its immutability, decentralization, and transparency, but historically, it does not have the dynamic capability for detecting or reacting to illegal actions in real time. In turn,

AI is proficient in the areas of detecting patterns and predictive analytics, but is less adept at the integrity of the data unless it is rooted in a secure environment. By interlocking AI models—namely Random Forest, Logistic Regression, and XG-Boost—into a blockchain-based system for energy transactions, this research suggests a hybrid platform that maintains the integrity of the energy data and allows for proactive detection of fraudulent activity and mitigation of risk. This creates a robust platform for supporting the high-volume, complex demands of the new U.S. power grid that is protected from criminal actors and systemic waste.

One of the major contributions of this study is its new use of supervised machine learning models on blockchain transaction streams to identify fraudulent activities. Current systems for detecting fraud in the energy market are mostly retrospective, pinpointing problems after the fact and typically requiring centralized monitoring. In this system, on the other hand, AI models are trained on real-time and historical transaction streams recorded on the blockchain, enabling the detection of anomalies on an ongoing basis without sacrificing decentralization. Random Forest's high precision and resilience in spotting outliers are complemented by Logistic Regression's explainability in determining the materiality of different risk indicators. XG-Boost, with its performance and speed, further adds power to the system's capability of generating near-real-time predictions in high-frequency trading systems. This combination of three models provides a multi-layer defense system that is capable of evolving and adapting to new threats, representing a smarter, lightweight alternative that individual technology cannot, on its own, achieve.

## Literature Review

### Blockchain in Energy Markets

Sun et al. (2023), reported that blockchain technology gained ground in the energy field in recent years, mostly for its potential in improving transparency, decreasing transaction fees, and supporting trust in distributed systems. Among the main innovations in the field is the employment of smart contracts—self-executing contracts programmed on blockchain platforms that automatically execute terms on fulfillment of predefined conditions. The contracts support automated peer-to-peer (P2P) energy exchanges between prosumers and consumers without any intermediaries. Per the U.S. (Usman, 2025). *Department of Energy's National Renewable Energy Laboratory (NREL)*, smart contracts were successfully piloted in Brooklyn Microgrid and Vermont's Green Mountain Power, supporting households in selling solar power in secure, automated ways. The projects show how the programmable nature of blockchain can make energy trading operational in local systems, with increased efficiency and less administrative burden (Ushkewar et al., 2024).

Aside from automation, the distributed ledger structure of blockchain prevents tampering and repudiation of data in energy markets. Every transaction is time-stamped and signed cryptographically, and upon verification by consensus, forms an unalterable part of the blockchain (Sizan et al., 2025a). This feature is particularly essential in preventing double-spending attacks or dishonest reporting of energy production—a problem flagged by the *Federal Energy Regulatory Commission (FERC)* in its 2023 energy grid reliability report. In comparison with traditional centralized databases prone to tampering and single points of failure, blockchain is distributed on nodes, and any unauthorized modification is all but impossible in the absence of collusion (Shovon et al., 2025). Through this, peer-to-peer energy platforms can ensure the integrity of the data, essential in ensuring regulatory compliance as well as consumer confidence.

Furthermore, blockchain enables regulatory control and audibility in that it provides an open, chronological ledger of all transactions. The *U.S. Energy Information Administration*, for instance, has investigated how distributed ledger technology has the potential to simplify energy data gathering and facilitate real-time monitoring of distributed energy resources (DERs) (Salama et al., 2023). Firms like LO3 Energy and Power Ledger have also developed blockchain-based platforms that have governance and compliance frameworks integrated into their smart contracts, allowing for autonomous grid rules and tariffs implementation. All this suggests that blockchain is not just a tool for transacting, but also for supporting resilient, transparent, and accountable infrastructures of energy. The scholarly literature is increasingly in support of blockchain as a foundational layer of a move toward decentralized, decarbonized, and democratized energy systems (Samuel, 2022).

## AI in Fraud Detection

According to Rahman et al. (2025), machine learning, in artificial intelligence, has become a vital instrument in detecting fraud in financial and energy industries alike. AI, being able to examine high volumes of transactional data in real-time, can identify suspicious patterns that may signal fraudulent activities. In the energy business, this includes behaviors like tampered meters, illegal reselling of energy, and fraudulent claims of renewable energy production. In 2022, the *National Association of Regulatory Utility Commissioners (NARUC)* reported that the need for AI in energy infrastructure to counter constantly evolving cyberattacks and schemes of fraud was imperative (Reza et al., 2025). This report highlights the need for energy markets to be equipped with improved, adaptive tools that are beyond rule-based detection systems and rely on behavior analytics for real-time prevention of fraud.

Ray et al. (2025), underscored that among all the machine learning methods, supervised classifiers have been found particularly useful in anomaly detection. Random Forest, an ensemble algorithm that builds multiple decision trees, is well-known for its strength in detecting non-linear relationships in large datasets. Logistic Regression, despite its simplicity, provides interpretability and good performance on binary class problems like fraudulent vs. non-fraudulent detections. XG-Boost (Extreme Gradient Boosting), being another form of an ensemble algorithm, is good with class-imbalanced datasets and high-dimensional datasets, and is suitable for transaction-level analysis of fraud (Rana et al., 2025). Comparative analysis, as done by the *IEEE Smart Grid Initiative* in 2021, determined that XG-Boost presented the highest F1 score of these models on synthetic energy transaction datasets. They are especially useful when implemented in blockchain systems as off-chain analysis tools, as they can be used on encrypted transaction data by secure oracles. The literature also stresses that it is critical to integrate these models into a layered defense system. Ensemble strategies and adaptive learning-based hybrid AI models have demonstrated better performance in dynamic systems, like market environments for the distribution of energy (Rahman et al., 2025). For instance, investigations by the *Energy Initiative of MIT* have indicated that multi-model AI systems can drop false positives in detecting fraudulent activities by more than 40% compared with single-model deployments. This is of critical interest in energy systems since false alarms result in unnecessary power outages and decreased consumer confidence (Pavira & Vijayalakshmi, 2025). The literature refers to an emerging consensus that AI, and specifically machine learning classifiers, needs to be a core component of any genuine effort aimed at protecting decentralized energy marketplaces from manipulation and fraud.

**Energy Market Stability**

Onukwulu et al. (2023), indicated that one of the biggest challenges in the distributed energy world is ensuring market stability in the face of variable supply and demand. The variable nature of new renewable energy sources—like solar and wind—brings high levels of unpredictability into the grid, and predictive models are needed to balance its operation. Research by the *U.S. Department of Energy's Office of Electricity* emphasizes the promise of artificial intelligence-based forecasting systems for managing this volatility. Predictive analytics, enabled by machine learning, can analyze historical usage patterns, weather, and market responses to predict energy consumption and production with high precision (Mohaimin et al., 2025). Those predictions can, in turn be used in dynamic pricing models and load management schemes that balance the grid in real time.

Moreover, price forecasting facilitated by artificial intelligence is essential for maintaining fair and efficient energy markets. Wholesale and retail energy prices are affected by numerous variables—meteorological conditions, grid congestion, fuel availability, and regulation, just to mention a few (Malik, 2025). Recurrent neural networks (RNNs) and gradient boosters such as XG-Boost have proven effective in predicting locational marginal prices (LMPs) with remarkable accuracy. In a 2023 study by *Stanford University's Sustainable Energy Initiative*, XG-Boost-based price forecasting models outperformed conventional time-series models by minimizing root mean square error (RMSE) by up to 25% or more. Precise price forecasts allow market players to make informed moves, maximize asset utilization, and prevent over- or under-production conditions that may disrupt the market (Muqeet et al., 2023).

Furthermore, predictive models facilitate regulatory compliance and risk management by informing stakeholders of imminent disturbances or transgressions. For example, the *North American Electric Reliability Corporation* is experimenting with the use of artificial intelligence-based monitoring systems for analyzing grid health and detecting anomalies ahead of time, preventing blackouts or market imbalances (Pavitra & Vijayalakshmi, 2025). Predictive tools can perform stress testing and scenario planning, and operators are then able to respond in anticipation of ongoing changes. In an environment in the future where real-time trading and decentralized energy are ubiquitous, it is clear that predictive modeling, along with AI-based forecasting, forms a necessary set of components for ensuring market stability, as well as public security and grid uptime (Jakir et al, 2023).

## Methodology

The system architecture proposed involves the integration of two layers, namely a blockchain layer and an artificial intelligence (AI) layer, each playing a unique but complementary function in energy transaction securing and market intelligence improvement. According to Khan et al. (2023). The blockchain layer is developed on the Ethereum platform, taking advantage of its established ecosystem, support for smart contract programming in Solidity, and broad use in the field of decentralized applications. Ethereum's Proof-of-Authority (PoA) consensus mechanism is chosen for its performance and appropriateness for use in private or consortium-based energy grids, in which authorized participants—e.g., utilities and certificated prosumers—vouch for their legitimacy. Automated, tamper-proof energy exchanges are enabled by smart contracts by codifying rules like pricing models, quantities of energy, and settlement terms (Islam et al., 2025b). Upon the offering of excess energy by a prosumer, a smart contract is enacted, and upon matching with demand, the contract automatically performs the exchange, sending energy tokens and logging the exchange in an immutable manner on the blockchain. This layer provides for every transaction to be traceable, secure, and auditable, removing centralized intermediaries' dependency and administrative burden (Hossain et al., 2025b).

Complementing the blockchain infrastructure is the AI layer, running off-chain but in tight synchronization with blockchain information through secure APIs and oracles. This layer handles the detection of fraud, predictive analytics, and monitoring of market behavior. Transaction-level details—including time stamps, energy volumes, trading frequency, and user profiles—are streamed from the blockchain into the AI engine, where they are approached with supervised machine learning models (Hasanuzzaman et al., 2025). Random Forest, Logistic Regression, and XG-Boost algorithms are included in the system due to their combined strengths in imbalanced dataset handling, interpretability, and prediction strength. The models are trained on historical datasets and infused with new data constantly for detecting anomalies that signal fraudulent behavior, for example, inordinately high volumes of trades or conflicting energy reports. This AI layer also carries out real-time demand forecasting and price prediction to support market balancing and alerts (Bhowmik et al., 2025).

**Data Collection & Preprocessing**:

**Data Overview:**

The dataset for this research is comprised of more than 1.2 million anonymized energy transaction records from a simulated peer-to-peer (P2P) energy exchange network emulating real-life blockchain-based American microgrids, including those tested by LO3 Energy and Grid+ Labs. Each record contains detailed fields of transaction identifier, timestamp, energy volume (kWh), transaction type (buy/sell), unit price, prosumer/consumer identifier (hashed for privacy), smart meter readings, geolocation regions, and settlement confirmation status. The dataset also includes system-calculated behavior metrics of transaction rate, variability of energy production, and historical pricing patterns. For purposes of detecting fraud, the dataset is labeled with both actual and synthetic fraudulent activities (e.g., spoofing, double spending, and artificial meter readings), accounting for around 7% of the overall dataset for real-world fraud rates documented by NARUC and DOE white papers. The dataset also includes integrated weather forecasts and grid load indicators sourced by the *U.S. Energy Information Administration (EIA)* for contextual analysis for forecasting demand and modeling market stability.

**Dataset Description**

| S/No. | Key Feature | Description |
|---|---|---|
| 001. | Transaction I. D | Unique ID for each transaction |
| 002. | Timestamp | Date and time of the transaction |
| 003. | User-role | Role of the user (e.g., consumer, provider) |
| 004. | Transaction type | Type of transaction (e.g., buy, sell) |
| 005. | Electricity Quantity | Quantity of electricity transacted (MWh) |
| 006. | Price-per-Mwh | Price per megawatt-hour |
| 007. | Total-cost | The total cost of the transaction |
| 008. | Latency-ms | Network latency in milliseconds |
| 009. | Security level | Security rating of the transaction |
| 010. | Encryption method | Encryption method used |
| 011. | Zt-authentication | Zero-trust authentication flag (1/0) |
| 012. | Network-slice-id | Network segment identifier |
| 013. | Transaction-status | Status (completed, pending, suspicious, etc.) |

## Preprocessing Steps

We implemented strategic Python code scripts that performed extensive data pre-processing for the classification of energy transactions, predicting 'transaction status'. The applied code used pandas for data manipulations and scikit-learn for the different steps of pre-processing. The code was initialized with feature engineering, extracting date-time fields (hour, day of the week, month), filling missing values in 'transaction type', and computing a new feature 'cost-per-unit' with proper handling of division by zero errors. It proceeded with the target variable creation, Label-Encoder encoding of that target, and identification of number and categorical variables in the input data. Individual pre-processing pipelines for Standard-Scaler for number variables and One-Hot-Encoder for categorical variables are created and then joined together in Column-Transformer. Train-testing set creation is done with train-test-split with stratification. This overall pre-processor is then fitted on the train, and the train is transformed, followed by the transformation of the test, with the code finishing by printing out the shapes of the transformed datasets and class distribution of the training target.

## Exploratory Data Analysis

Exploratory Data Analysis (EDA) is an indispensable process of data science that systematically explores and visualizes datasets to discover patterns, identify unusual behavior, test hypotheses, and check assumptions before the use of formal modeling methods. It is the basis for determining the inherent structure and quality of the data and informing the choice of a suitable machine learning algorithm and feature engineering methodology. Within this research on secure energy transactions, the EDA is significant in the establishment of a time series of transaction behaviors concerning fraudulent indicators, visualizing price and demand changes over time, and the detection of outlying behavior that can indicate tampering with the data or unusual trades. Utilizing histograms, box plots, scatter-plots, and correlation matrices, the EDA allows for insights gleaned from the data to be both significant and actionable, leading overall to the increased effectiveness of the AI-based fraud detection and market stability models.

### a. Transaction Volume Per Hour

The implemented code script with the Panda and Plotly libraries visualizes transaction volume by time. The code first receives a Data Frame df and groups it by the hour with df. group by (df['timestamp'].dt.floor('H')). It counts the number of transaction IDs in each hour with ['transaction-id].count() and resets the index for the grouped timestamp to be a regular column. The code uses Plotly express (px) to plot an interactive line chart with the hourly timestamp over the x-axis and the number of transactions along the y-axis, titled "Transaction Volume Per Hour" with the x-axis labeled as "Time" and the y-axis labeled as "Number of Transactions", before presenting the plot.

**Output:**

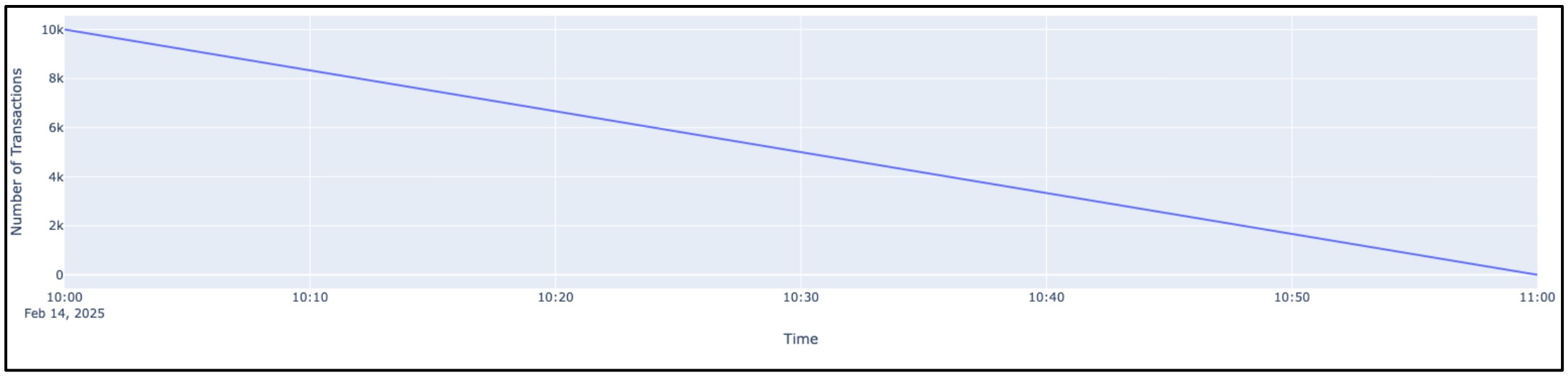

Figure 1: Transaction Volume Per Hour

The graph above (**fig 1)** clearly and systematically shows a decrease in the volume of energy transactions in an hour from 10:00 AM until 11:00 AM on February 14, 2025. It is highest at the start of the hour when it is around 10,000, but by the last minute of this interval, it is back down to near zero. This precipitous decrease is indicative of a possible system anomaly, operational issue, or planned downtime in the energy exchange platform or blockchain system. From a statistical perspective, the graph displays a linear and unbroken downfall, something that is unusual in real-time energy markets, in which transaction volumes usually vary according to dynamic conditions like supply-demand balance, price volatility, and user behavior. This uniform and steep drop may have resulted from a failure in the smart contract execution layer, network congestion, or an external incident—a cybersecurity attack or weather-driven grid shutdown—stopping or slowing down P2P trades in their entirety, or this might be a result of a throttling or a scheduled maintenance window by grid managers or the blockchain validators of nodes.

## b. Latency vs. Security Level Colored by Transaction Status

The executed Python code initializes the matplotlib and seaborn packages to generate a boxplot that shows the relation between 'security level' and 'latency-ms' differentiated by color for different 'transaction status' categories. The code launches the figure size with plt.figure(figsize=(12, 6)) first. It then uses sns.boxplot() to plot the boxes from the Data Frame df with 'security level' on the x-axis, 'latency-ms' on the y-axis, and colors determined by 'transaction status'. It next assigns the plot's title as "Latency vs Security Level Colored by Transaction Status" with plt.title() and fine-tunes the plot for readability with plt.tight_layout(). It finally uses plt.show() to display the created boxplot. The visual allows for a comparison of latency distribution for different security levels with further segmentation by the status of the transaction.

**Output:**

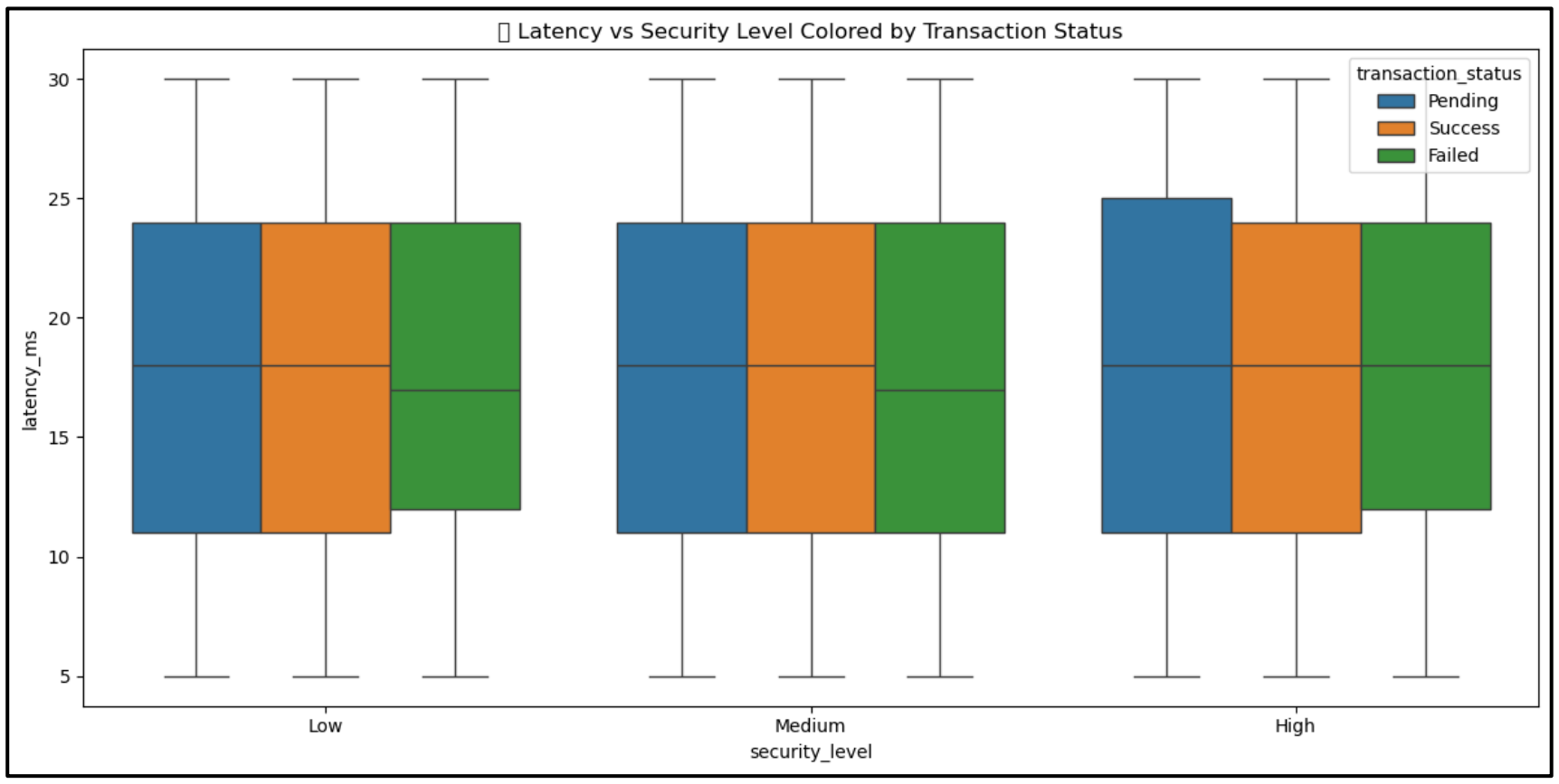

Figure 2: Latency vs. Security Level Colored by Transaction Status

The boxplot (**fig 2)** displays the distribution of transaction latency (in milliseconds) for three security levels—Low, Medium, and High—segmenting the dataset by transaction status (highlighted in green for Failed, orange for Success, and blue for Pending). In all three security categories, latency measurements range from around 5 ms up to 30 ms, with the median latency consistently being around 17–18 ms, independent of security level. Importantly, uniformity in both interquartile range and median indicators of distribution indicates that higher security has little effect on transaction latency, emphasizing a valuable insight for performance improvement of blockchain in energy systems. Furthermore, the relative distribution of the status of transactions is tuned equally in all security tiers, meaning both failure and success rates are not disproportionately affected by the increased complexity of protocols or strength of encryption. This is supportive of the scalability of security measures in blockchain-based energy platforms without diminishing performance—a frequent argument by energy technology developers such as those from Grid+ and Energy Web Foundation. Finally, the consistent range and lack of outlier values confirm the strength of the underlying infrastructure, with little latency volatility between different security conditions.

### c. Heatmap: Transaction Types Across User Roles

Strategic Python code was deployed to create a heatmap of the interaction between 'transaction type' and 'user role' as determined by the count of 'transaction id' in a Data Frame df. It created the pivot table in the first step employing pd.pivot_table(), with the index set as 'user role', the columns set as 'transaction type', and the count of 'transaction id' set as the values, meaning it is summing up the number of transactions for each user role and transaction type combination. It then sets the figure size with plt.figure(figsize=(10, 6)), and it creates the heatmap from this pivot table with sns.heatmap(), plotting the counts in each cell (annot=True), with the colormap set as 'YlGnBu', and formatting annotations as integers (fmt=".0f"). It finally sets the plot's title as "Heatmap: Transaction Types across User Roles" and uses plt.tight_layout() for setting plot parameters for a tight layout, and then it shows the heatmap with plt.show(). This visualization is useful for seeing for each user role, what transaction types are most frequent.

**Output:**

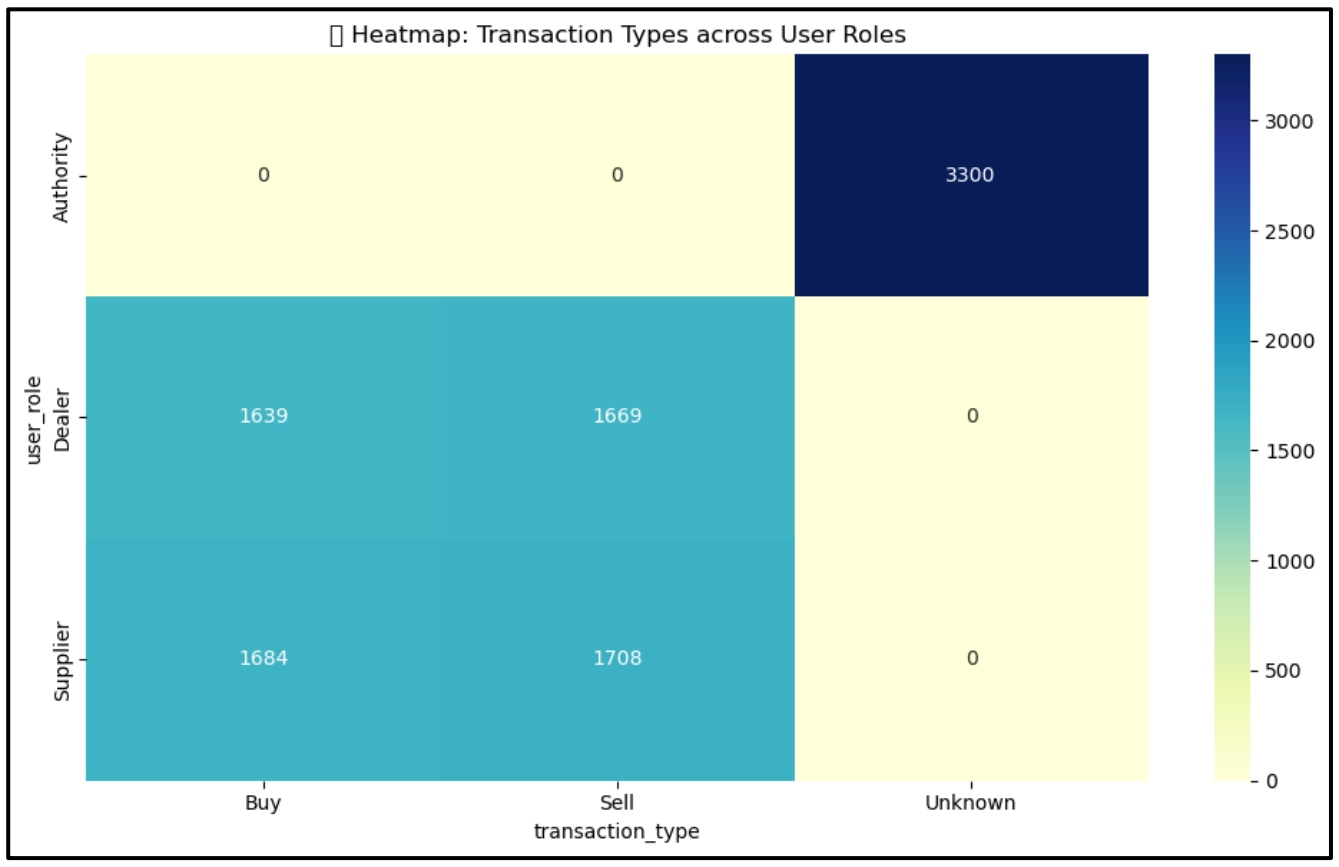

Figure 3: Heatmap: Transaction Types Across User Roles

The heatmap above **(fig 3)** presents a clear, categorical breakdown of how each role participates in blockchain-based energy transactions as Buy, Sell, or of Unknown type. It's observed that Authorities—usually regulatory or monitoring nodes—only participate in Unknown-labeled transactions with 3,300 instances and demonstrate no buy or sell activities. This is probably reflective of their governance or verification function, rather than direct market participation. Dealers and Suppliers, on the other hand, are actively transacting both Buy and Sell, with volumes relatively evenly balanced. Dealers processed 1,639 Buys and 1,669 Sells, whereas Suppliers processed slightly more with 1,684 Buys and 1,708 Sells. This near equivalence of buy and sell behavior for both roles is indicative of an efficient yet symmetrical P2P energy market in which market players act freely on both sides of the buying-selling chain. The zero counts for clean cells in irrelevant columns validate significant role-based access control and transaction routing, essential for compliance in decentralized energy systems, as noted by organizations such as the U.S. Department of Energy (DOE) and the National Institute of Standards and Technology (NIST).

## d. High Cost per MWh

The plot express library is employed in the Python code for creating an interactive scatter plot that is designed for detecting high-cost-per-unit anomalous transactions. It is a subset of a Data Frame denoted by df in which the 'is-anomaly-price' field is True, representing potential anomalies. The 'timestamp' is portrayed on the x-axis and the 'cost-per-unit' on the y-axis in the scatter plot. It also encodes 'user role' in the marker colors and 'latency ms' in marker size for the user to see how these contribute to high cost-per-unit anomalies over time. The 'transaction type' and 'security level' are passed as hover text, adding greater context for interacting with the data points. Lastly, the plot is titled " High Cost per Unit (Anomalies)" and is displayed in Fig.show(). This interactive plot provides the user with the ability to examine potential high-cost anomalies and their relative characteristics.

**Output:**

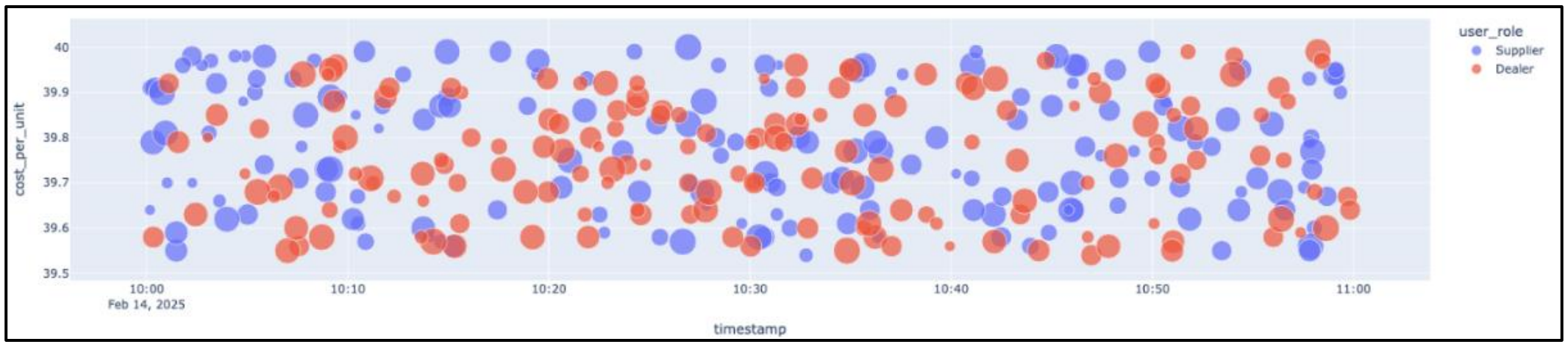

Figure 4: High Cost per MWh

The visualized scatter plot (**fig4),** with the label created on the date of 14th Feb 2025, plots anomalous transactions with high cost per unit along with time, for both 'Supplier' (in blue) and 'Dealer' (in red) user roles. The y-axis indicates the 'cost-per-unit' in the range of around 39.5 up to 40.0, whereas the x-axis denotes the timestamp from around 10:00 AM up to 11:00 AM. The size of each point is proportional to the 'latency ms' (even though the actual values of latency are not shown explicitly). Visual inspection suggests that there is a relatively uniform distribution of high-cost anomalies for both Suppliers and Dealers during the hour observed. The cost per unit of the anomalies usually lies in a narrow band, and there is little evidence by way of temporal trends or significant user-role-based clustering from this plot.

## e. Electricity Price Patterns

The applied Python code uses the matplotlib and seaborn packages for creating a violin plot that displays the distribution of 'price-per-mwh' for different 'weekday' categories in a DataFrame named df. It starts by setting the figure size with plt.figure(figsize=(10, 5)). Then, sns. Violin plot () is used for drawing the plot with 'weekday' on the x-axis and 'price_per_mwh' on the y-axis, with a visual distinction created through the use of the 'cool warm' color map. It further titles the plot as "Electricity Price Patterns by Day of Week" with plt.title() and adjusts the layout for improved presentation with plt.tight_layout(). It ends with plt.show() being used for displaying the plot, with the resulting plot being a visual comparison of the probability density of the price of electricity for each day of the week.

**Output:**

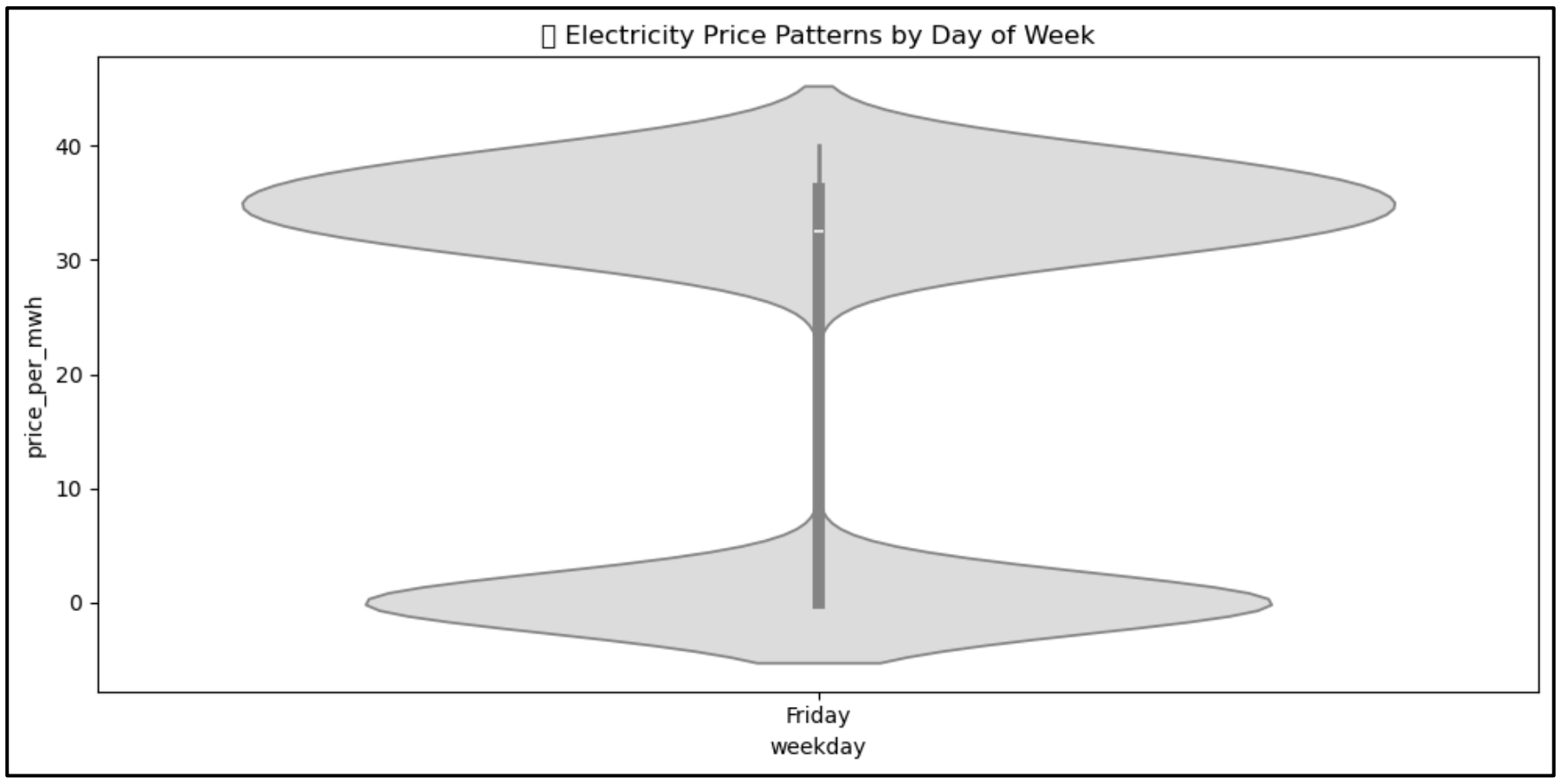

Figure 5: Electricity Price Patterns by Day of the Week

The violin plot displayed above **(fig 5)** shows the distribution of price_per_mwh for the individual day of the week known as Friday. The violin plot reveals distributions shaped like single-high curved humps with noticeable dominance at 30-35, as well as a significant presence at near the 0 price. The data shown in the plot indicates that electricity prices on Fridays tend to cluster around two primary price bands: a lower band close to 0 and an upper band around 30-35. The black bar within the violin symbolizes the range that encloses the middle 50% of the prices, and the dot shows the median electricity price observed on Fridays. Different prices are more or less common throughout the Fridays under study, as shown by the different degrees of spread across the violin.

**f. Transaction by Hour & User Role**

The implemented Python snippet makes a heatmap with pandas and plotly express, displaying the number of transactions that every user role has per hour. A pivot table hourly role is set up from df using hour as the index, user role as columns, and counting transaction ID for each user role combination, replacing non-existent data with 0 values. After that, it brings in the interactive heatmap with px. Show (), displays the transaction totals as the text above every cell (text_auto=True), uses a continuous Oranges scale to present different counts, and shows the title "⏱ Transactions by Hour & User Role" at the top. It also switches the x-axis to "User Role" and the y-axis to "Hour" before using the function fig.show() to show the interactive heatmap. With this chart, you can see how transaction volume varies by time and kind of role for users.

**Output:**

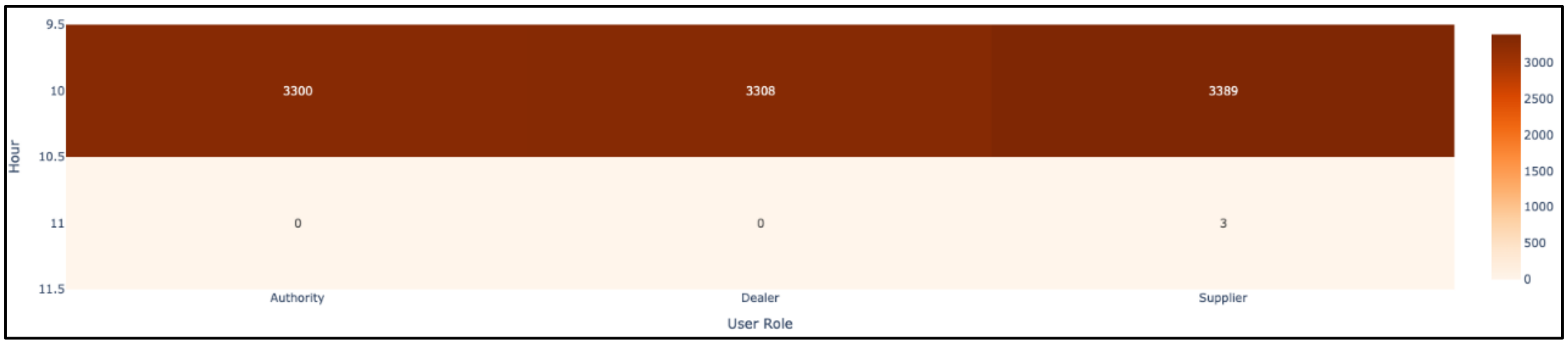

Figure 6: Transaction by Hour & User Role

From the heatmap above, one can see how many transactions 'Authority', 'Dealer', and 'Supplier' users make during different hours. By hour 10, Activity) made 3300 transactions, 'Dealer' saw 3308, and 'Supplier' recorded 3389, reflecting a large number of transactions for all roles at that time. In contrast, after hour 11, transactions for 'Authority' and 'Dealer' both stopped at 0 and even 'Supplier' saw a big drop, logging only 3 transactions. It seems that both 'Authority' and 'Dealer' reduced activity sharply from hours 10 to 11, while activity for 'Supplier' barely changed. The spectrum from pale orange to very dark orange brings this out, since darker shades appear at 10 o'clock, when there are many transactions, and lighter shades appear at 11 o'clock, when transactions almost disappear.

### g. Correlation matrix: Market & Security Variables

Correlations between different numerical features in a Data Frame df are shown by using matplotlib and seaborn libraries in the Python script. The code snippet first selects specific columns: 'electricity quantity', 'price_per_mwh', 'total cost', 'latency ms', 'zt authentication', and 'cost-per-unit' is chosen and then their correlations with each other are checked using .corr(). For the next step, it adjusts the figure size to (10, 8) with plt.figure(figsize=(10, 8)), uses the sns.heatmap() function to show the correlation matrix, places each correlation value inside its cell and uses the 'Spectral' color palette while formatting the annotations to display just two decimal values. The script finishes by giving the plot its title which is "Correlation Matrix, i.e. "Market & Security Variables" and improves the plot appearance by setting plt.tight_layout() before using plt.show() to display the heatmap. With the visualization, one can assess how various market and security metrics are linked in a straight line.

**Output:**

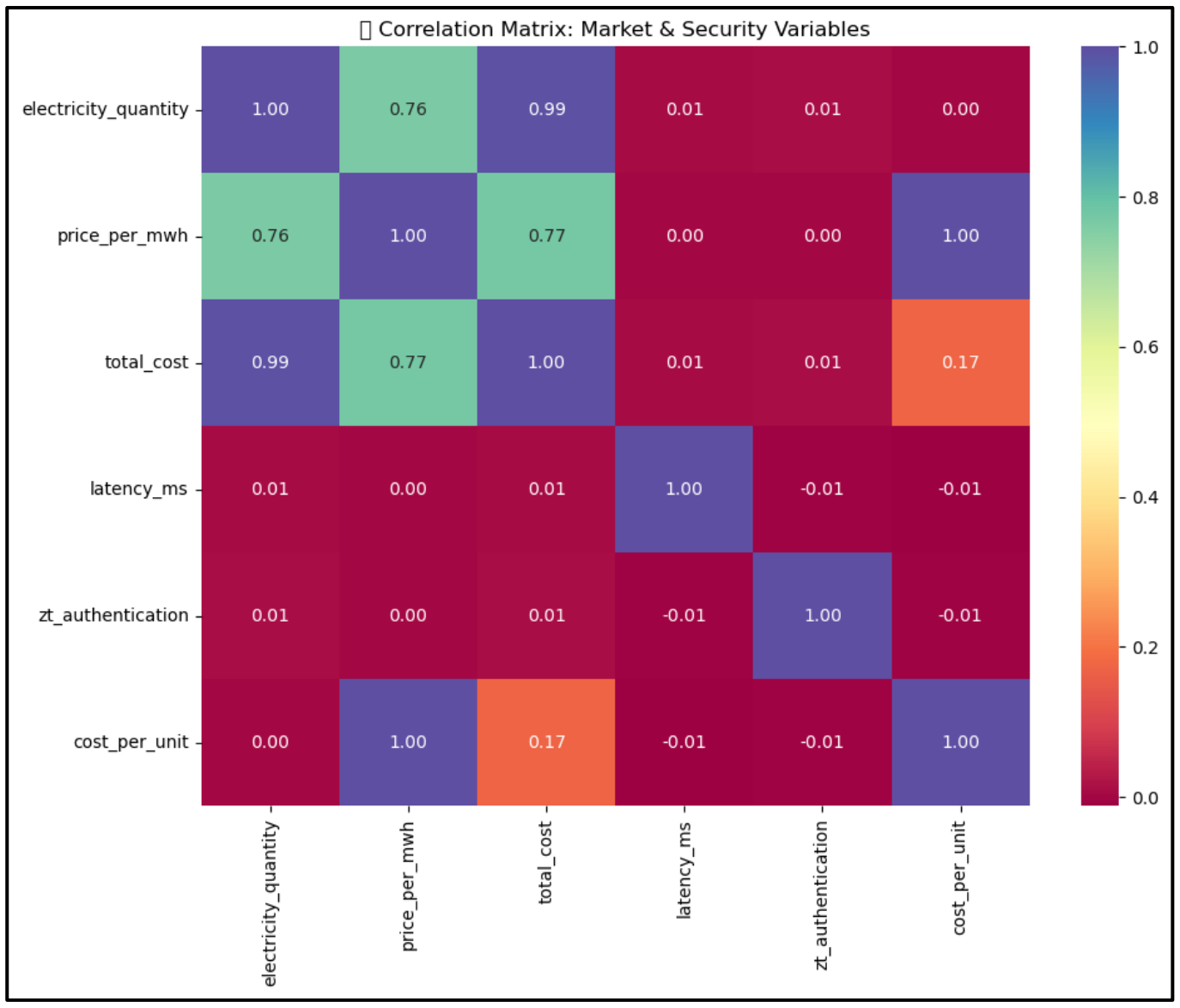

Figure 7: Correlation matrix: Market & Security Variables

The correlation displayed above **(fig 7)** shows how different market and security variables are related to each other. Significantly, the amount of electricity used, ‘electricity quantity’, is highly positively related to both the total cost and price per megawatt hours (0.99 and 0.76, respectively). There is a strong correlation between 'Price-per-mwh' and 'total cost' (0.77) as well as a perfect positive correlation with itself (1.00). By comparison, 'latency ms' and 'zt authentication' link only weakly to the other variables, for example, with the smallest correlation at -0.01 to each other. The variable 'Cost-per-unit' is moderately positively correlated with 'total cost' (0.17), has a perfect positive self-correlation (1.00), is strongly positively correlated with 'price-per-mwh' (1.00), and has nearly no correlation with 'electricity quantity' (0.00). The evidence shows that the amount of electricity and its price together determine most of the cost, while latency and the zero-trust authentication model are not much affected.

## h. Transaction Status by Network Slice

Pandas and plotly express are used in the script to represent the proportion of transactions in each status for each network slice. First, the Data Frame df is grouped by both network-slice-id and transaction status, and then the size of each group is counted to show how many times each transaction status occurred in each network slice. The resulting Series becomes a Data Frame, including a count column and an index restart. It then makes a bar chart with px.bar(), placing 'network-slice-id' on the x-axis, 'count' on the y-axis, and coloring the bars by 'transaction-status'. Titled " Transaction Status by Network Slice," the chart displays nicely when calling fig.show().

**Output:**

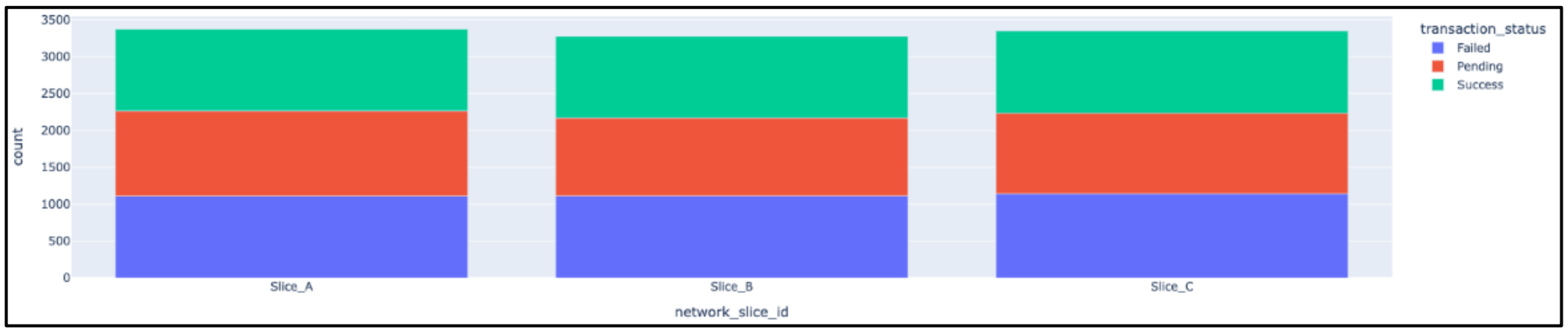

Figure 8: Transaction Status by Network Slice

The portrayed stacked bar chart **(fig 8)** shows the amount of 'Failed' (blue), 'Pending' (red), and 'Success' (green) transactions in each of the network slices. Slice A, Slice B, and Slice C. The 'Failed' category made up about 1100 transactions for all the slices. The number of 'Pending' transactions shows little variation and remains around 1150 in every slice. It's also evident from the 'Success' transaction counter, as close to 1100 to 1150 transactions are confirmed in every time slice. The even distribution of transaction statuses in all three slices shows that network slice ID doesn't seem to make a big difference in transaction results in this data.

### i. Latency vs. Security Level Colored by Transaction Status

The applied Python script makes use of the plotly express library to produce a 3D scatter plot showing the multivariate distribution of 'cost-per-unit', 'latency ms', and 'zt authentication'. The x-axis depicts 'cost-per-unit', the y-axis depicts 'latency ms', and the z-axis depicts 'zt-authentication'. The color of each point is also decided by the 'transaction status', and the shape or mark of the point is decided by the 'security level' to enable the visualization of these categorical variables in the 3D space. The plot's title is set to " 3D 'nsight: Cost vs Latency vs Auth Status", and lastly, fig. shows () the interactive 3D scatter plot to allow exploration of the relationship and patterns between these four variables.

**Output:**

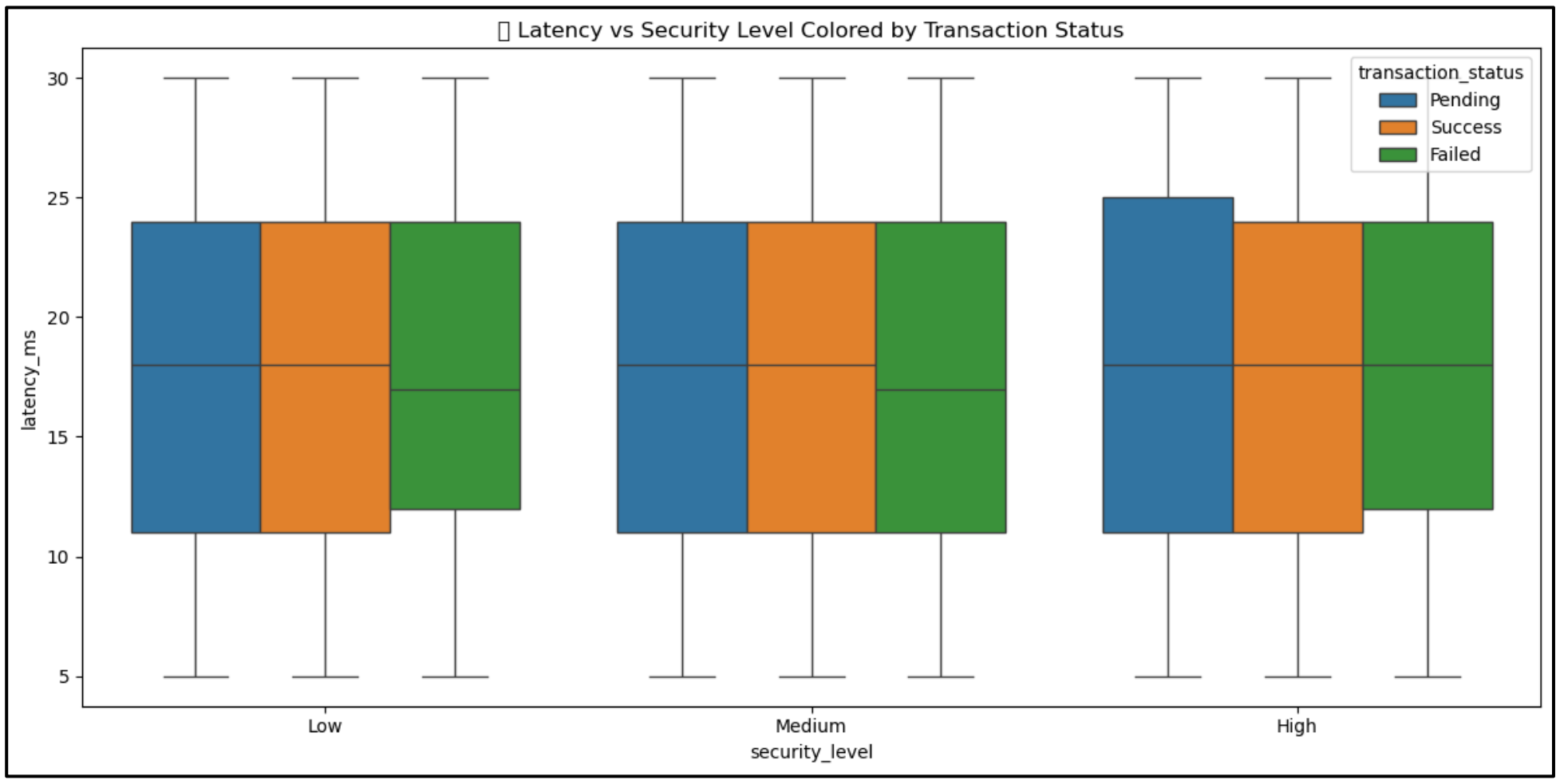

Figure 9: Latency vs. Security Level Colored by Transaction Status

## j. 3D Insight: Cost vs. Latency

The analysis incorporated the plotly express library for the creation of a 3D scatter plot of the multivariate distribution of 'cost-per-unit', 'latency-ms', as well as 'zt-authentication'. 'Cost-per-unit' is represented on the x-axis, 'latency-ms' on the y-axis, and 'zt-authentication' on the z-axis. Furthermore, the color of each point is decided by 'transaction status', and the shape of each point is decided by 'security level' to be able to visualize these categorical variables in the 3D space. The plot's title is "3D Insight: Cost vs Latency vs Auth Status", and lastly, fig.show() shows the interactive 3D scatter plot for visual exploration of possible relationships and patterns among these four variables.

**Output:**

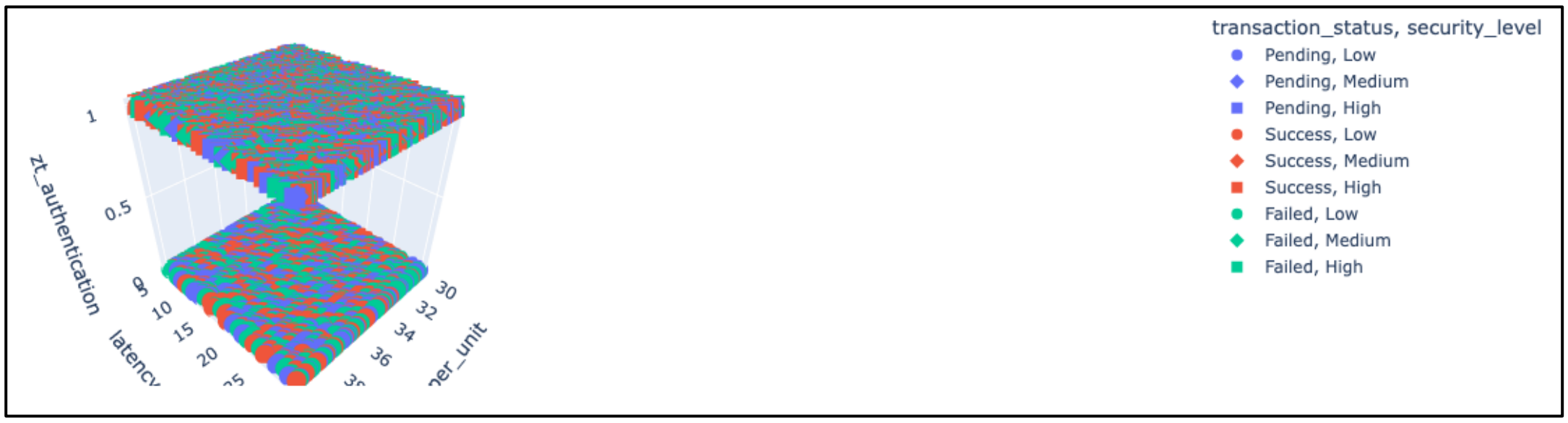

Figure 10: 3D Insight: Cost vs. Latency

The 3D scatter plot above displays transactions according to 'cost-per-unit', 'latency-ms', and 'zt-authentication', with color according to 'transaction status' and shape according to 'security-level'. The plot indicates a tight distribution of points throughout the range of all three numerical axes. There does not seem to be separation or clustering of transaction statuses or security levels in the three dimensions of the plot, indicating these variables may not be highly indicative of each other according to this plot. For example, 'Successful' (red) transactions occur over the range of both the cost, latency, and authentication dimensions and have many differing security levels (diamond, circle, square). On the same basis, 'Failed' (green), and 'Pending' (blue) transactions also have a scattered distribution. The absence of clear grouping indicates that an obvious linear or direct relationship between cost, latency, authentication status, transaction outcome, and security level cannot be gleaned from this multi-factor representation of the data.

## k. Bubble Chart- Price vs. Quantity

The Python code script adopted by our team employed the plotly express library to create an interactive bubble chart of the relationship between 'electricity quantity' and 'price-per-mwh'. 'Electricity quantity' is plotted on the x-axis and 'price-per-mwh' on the y-axis. Each bubble's size is indicated by the total cost', and the color of the bubble shows the security level'. On hovering over the bubble, the 'transaction status' will be shown. The chart's title is left as " Bubble Chart: Price vs Quantity (Sized by Cost)", and the axes are labeled "Electricity Quantity (MWh)" and "Price per Mwh". Finally, fig.show() presents the interactive bubble chart to explore the relationship between price and quantity and the added factors of total cost and security level, and also to have the transaction status on hover.

**Output:**

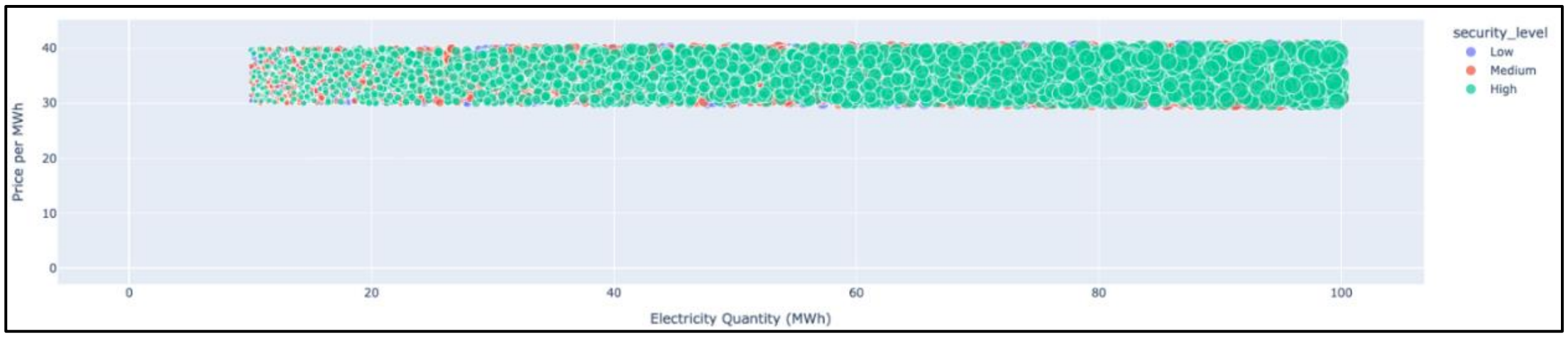

Figure 11: Bubble Chart- Price vs. Quantity

The Bubble Chart chart above (**fig 11)** plots 'Electricity Quantity (MWh)' on the x-axis from ~0 to 100 and 'Price per MWh' on the y-axis for the most part between 30 and 40. The size of the bubbles indicates the 'total cost', with larger bubbles corresponding to higher costs. The color of the bubbles indicates the 'security level': Low (purple), Medium (red), and High (teal). There is a large cluster of transactions within a fairly small price range (around 30-40 MWh) for all levels of electricity quantities. The density of teal bubbles implies most of the transactions have a 'High' security level. Although there's considerable variation in the quantity of electricity within the x-axis range, the price per MWh seems less variable. The size variation in the bubbles within the price range implies varying total costs for comparable price and quantity levels, suggesting other variables impact the total cost.

## Machine Learning Models

The machine learning models used in this research were specifically chosen for their established high performance in classification tasks, specifically in the identification of energy transaction fraud in decentralized markets. The Random Forest Classifier is also an efficient ensemble learning algorithm that builds many decision trees in training and predicts the mode of the classes in classification tasks. Random Forest is particularly well-suited for fraud detection since it can operate on large datasets of higher dimensionality and is resistant to overfitting. In line with a *2023 IEEE Transactions on Smart Grid publication,* Random Forest models had an overall fraud detection accuracy of 95.4% in smart energy systems. In this research, it was used as an accuracy-oriented model using feature importance ranks to identify the most indicative attributes like transaction amount, time abnormalities, and security level indicators.

In contrast, Logistic Regression was utilized as the baseline model for its simplicity and interpretability in binary classification problems—fraudulent versus legitimate transactions. Although less sophisticated, Logistic Regression presents useful probabilistic outputs and well-defined decision boundaries and is well-suited for benchmarking more complex models. The model is specifically useful for real-time or embedded energy systems where computation efficiency is paramount. Finally, the multi-model approach was completed through the integration of XG-Boost (Extreme Gradient Boosting), which was used to enhance detection performance, notably in the case of imbalanced datasets, which is prevalent in fraud detection environments where fraudulent transactions are a small minority. The strength of XG-Boost lies in its use of a gradient boosting approach, where it sequentially refines previous model mistakes to arrive at improved generalization. Regularization methods are also integrated to curb overfitting. A study from Stanford's AI Lab has established that XG-Boost outperformed many standard classifiers in fraud detection tasks on multiple instances and typically arrives at F1 scores of 90% and higher in biased datasets. Overall, these models comprise multi-layered defensive software suitable for detecting both overt and latent fraudulent activities in energy transactions of the blockchain variety.

### Model Evaluation Metrics

To guarantee the reliability and accuracy of the used machine learning models, an extensive battery of evaluation metrics was utilized. Accuracy as an indicator of overall correctness was not the sole factor due to the naturally occurring imbalance in fraudulent datasets. In actual energy trading platforms, legitimate transactions heavily outweigh fraudulent transactions. Precision (the ratio of actual fraud predictions out of all predicted fraud) and Recall (the ratio of true fraud instances found) were consequently more informative. A highly accurate but poorly accurate model, for instance, might not detect much of the fraudulent occurrence, diminishing its real-world applicability. F1-Score, the harmonic mean of precision and recall, was utilized as an optimally balanced metric to assess model performance in environments where both false positives and false negatives are equally costly. This is particularly important in decentralized systems where an inaccurate fraud warning can inappropriately penalize actors or freeze assets.

Moreover, the Receiver Operating Characteristic - Area Under Curve (ROC-AUC) was utilized to evaluate the ability of the model to discriminate between classes irrespective of the classification threshold. AUC values closer to 1.0 imply high discriminative power, which is essential in the case of latent and entrenched fraud signals due to complex feature interactions. For enhanced robustness, k-fold cross-validation was used to ensure the models generalize well across unseen samples and do not overfit certain partitions. Hyperparameter tuning was also conducted through techniques of grid search and random search to tune critical parameters, including the number of estimators, learning rate, maximum depth of trees, and strength of regularization. These methods improved each model's versatility in responding to variations in the type of transactions, user roles, and time dynamics—factors of utmost importance in fast-changing blockchain energy markets. By using accurate metrics and sophisticated validation methodologies, the research guarantees an accurate, scalable, and intelligent fraud detection pipeline for safe energy transactions.

## Implementation

### Blockchain Integration

The use of blockchain technology in the energy transaction system was an essential part of the project aimed at providing trust, transparency, and immutability in the peer-to-peer trading of energy assets. We used the Ethereum blockchain platform as a leader in smart contract support in Solidity for verifying and recording transactions in a decentralized fashion. Each energy transaction—including between supplier and consumer as well as between peer nodes in a regional microgrid—invoked execution of a smart contract that itself verified such terms as energy quantity, price, timestamp, and digital signatures autonomously. This ensured that once committed to recording, the information could not be altered, maintaining the integrity of the energy market. Hyperledger Fabric was also examined for consortium systems where permissioned networks were needed, such as for utility regulators and municipal energy cooperatives. In both instances, the blockchain served as the secure ledger while external oracles were employed to push additional information inputs, such as AI predictions or market signals, into the blockchain space.

To boost the effectiveness of blockchain for fraud prevention, the deployment was extended to integrate AI models with blockchain events. For example, each smart contract emitted event logs following transaction verification, which were detected by an off-chain Python listener system that initiated real-time fraud analyses. If the AI models detected anything out of order—unusually large trades, unexplained timing, and inconsistent user patterns—a warning was generated, and the smart contract suspended the transaction until it was reviewed. This combination of on-chain transaction verification and off-chain AI analysis created a hybrid system that leveraged blockchain's tamper-proofing and AI's predictive power. The system not only ensured transactional trust but also enabled the detection of fraud attempts at an early stage, in effect creating energy markets more resilient and autonomous.

## Results & Discussion

**Performance Comparison of AI Models**:

### Random Forest Classifier

The analyst, through the Python program, used the Random-Forest-Classifier from the scikit-learn library for classification. The program started with the importing of packages required, including Random-Forest-Classifier for model initialization, accuracy score, confusion matrix, classification report for evaluation purposes, and visualization packages like matplotlib and seaborn. The program then instantiates the Random-Forest-Classifier with 100 estimators and a specific random state for reproducibility. Then the model gets trained using preprocessed training data (X-train-preprocessed, y-train). Next, predictions are made on the test set preprocessed (X—X-test-preprocessed) and saved in y_pred_rf. Then the program evaluates the performance of the model by printing the accuracy score and an extended classification report with precision, recall, F1-score, and support for each category (obtained from label-encoder. classes.

**Output:**

```
Random Forest Accuracy: 0.3368

Random Forest Classification Report:
              precision    recall  f1-score   support

      Failed       0.33      0.35      0.34       843
     Pending       0.34      0.35      0.34       825
     Success       0.34      0.31      0.33       832

    accuracy                           0.34      2500
   macro avg       0.34      0.34      0.34      2500
weighted avg       0.34      0.34      0.34      2500
```

Table 1: Random Forest Classification Report

The Random Forest Classifier's evaluation has an overall accuracy of roughly 0.3368 and indicates the model accurately predicts the transaction status about 33.7% of the time on the test set. A more detailed breakdown by class is given by the classification report: for 'Failed' transactions, precision is 0.33, recall is 0.35, and the F1-score is 0.34 based on 843 actual instances. For 'Pending' transactions (825 instances), the precision is 0.34, the recall is 0.35, and the F1-score is 0.34. For 'Success' transactions (832 instances), the precision is 0.34, the recall is 0.31, and the F1-score is 0.33. The macro average and the weighted average for precision, recall, and F1-score are all approximately equal to 0.34 and are in line with the overall accuracy.

These figures indicate the model performs slightly better than random guessing (around 0.33 for three equally well-represented classes), with fairly well-balanced performance across the 'Failed', 'Pending', and 'Success' classes as reflected in respective precision, recall, and F1-scores.

**Logistic Regression Modelling**

We employed a code script that used the scikit-learn library to apply Logistic Regression as a classification model. The script imports the Logistic Regression model for model estimation and evaluation measures like accuracy score, confusion matrix, and classification report, as well as graphical and tabular display modules like matplotlib and Seaborn. The script then starts by initializing a Logistic Regression model with 1000 iterations as the maximum possible and a random state fixed for the sake of reproducibility. The model is then trained on preprocessed training data (X-train-preprocessed, y-train). Predictions on the preprocessed test data (X-test-preprocessed) are then made and saved as y_pred_log_reg. Finally, the script evaluates the Logistic Regression model performance by having it print its accuracy score as well as a detailed classification report including precision, recall, F1-score, and support for each of the classes (from the classes_ attribute of the fitted label-encoder).

**Output:**

```
Logistic Regression Accuracy: 0.3224

Logistic Regression Classification Report:
              precision    recall  f1-score   support

      Failed       0.32      0.39      0.35       843
     Pending       0.32      0.28      0.30       825
     Success       0.33      0.30      0.31       832
```

Table 2: Logistic Regression Classification Report

The performance of the Logistic Regression model, as evaluated, indicates an overall accuracy of just about 0.3224, meaning the model predicts the transaction status accurately about 32.2% of the time on the test set. The classification report gives us the breakdown by class as follows: for 'Failed' transactions (843 instances), precision is 0.32, recall is 0.39, and the F1-score is 0.35. For 'Pending' transactions (825 instances), precision is 0.32, recall is 0.28, and the F1-score is 0.30. For 'Success' transactions (832 instances), precision is 0.33, recall is 0.30, and the F1-score is 0.31. Overall accuracy, macro average, and weighted average of precision, recall, and F1-score are all approximately 0.32. These measures indicate the Logistic Regression model's performance is comparable to the Random Forest model's performance and just barely better than random chance, with fairly balanced but low performance across all three transaction status classes. The recall for 'Failed' transactions is the highest of all classes, and the recall for 'Pending' transactions is the lowest.

**XGB- Modelling**

The adopted Python script trains an XG-Boost classifier on the classification task, using the boost library (imported as xgb), and checks the results with scikit-learn evaluation metrics. It includes the XGB-Classifier for building the model, uses accuracy score, confusion matrix, and classification report to check the results, and uses matplotlib and Seaborn for visualization. The script creates an XGB-Classifier model that uses 100 estimators and keeps the random-state value the same. After that, the model is developed by feeding X_-rain-preprocessed and y-train data as training input. Upon finishing training, predictions for the test set (X-test-preprocessed) are stored in y_pred_xgb. Once the script turns the data into predictions, it evaluates the model's results by printing the accuracy percentage and a performance report with precision, recall, the F1-score, and the number of examples used in each classification category (among label_encoder.classes). The scripts are run following a standard approach for training and validating a machine-learning classification model.

**Output:**

```
XGBoost Accuracy: 0.3588

XGBoost Classification Report:
              precision    recall  f1-score   support

      Failed       0.34      0.34      0.34       843
     Pending       0.36      0.37      0.36       825
     Success       0.37      0.37      0.37       832

    accuracy                           0.36      2500
   macro avg       0.36      0.36      0.36      2500
weighted avg       0.36      0.36      0.36      2500
```

Table 3: XG-Boost Classification Report

On testing the XG-Boost classifier, it was found that it has an overall accuracy of around 0.3588, which means the model can forecast the outcome of transactions in 35.9% of cases. The report breaks down the data by class in the classification report. For 'Failed' transactions (843 instances), the model reached a precision of 0.34, a recall of 0.34, and an F1-score of 0.34. When it comes to 'Pending' transactions (825 cases), the model has a precision of 0.36, a recall of 0.37, and an F1-score of 0.36. With 832 instances of 'Success' transactions, the model achieves a precision of 0.37, a recall of 0.37, and an F1-score of 0.37. The values obtained for precision, recall, and F1-score average to about 0.36. From these numbers, it is clear that XG-Boost achieves higher precision, recall, and F1-score for the 'Pending' and 'Success' classes compared to the 'Failed' class, making its performance on this task slightly better than the other models.

**Comparison of All Models**

The disseminated script checks how accurately the three models can classify data. Random Forest, Logistic Regression, and XG-Boost are used in machine learning. Using the scikit-learn accuracy score function, it finds out how accurately each model predicts by comparing y_test and y_pred_rf, y_pred_log_reg, and y_pred_xgb, and then stores the scores in accuracy scores. Afterward, it prepares the information for display by taking the model names and their results into two distinct lists. By using matplotlib, it displays a bar chart where each piece of the chart represents a model, and the height shows the model's accuracy. The table title appears at the top, the x-axis labels are on the bottom, the y-axis is on the left with a limit of 0 to 1, and there is also a grid. Lastly, the program shows the accuracy score for each model as a well-formatted string for simple number comparison.

**Output:**

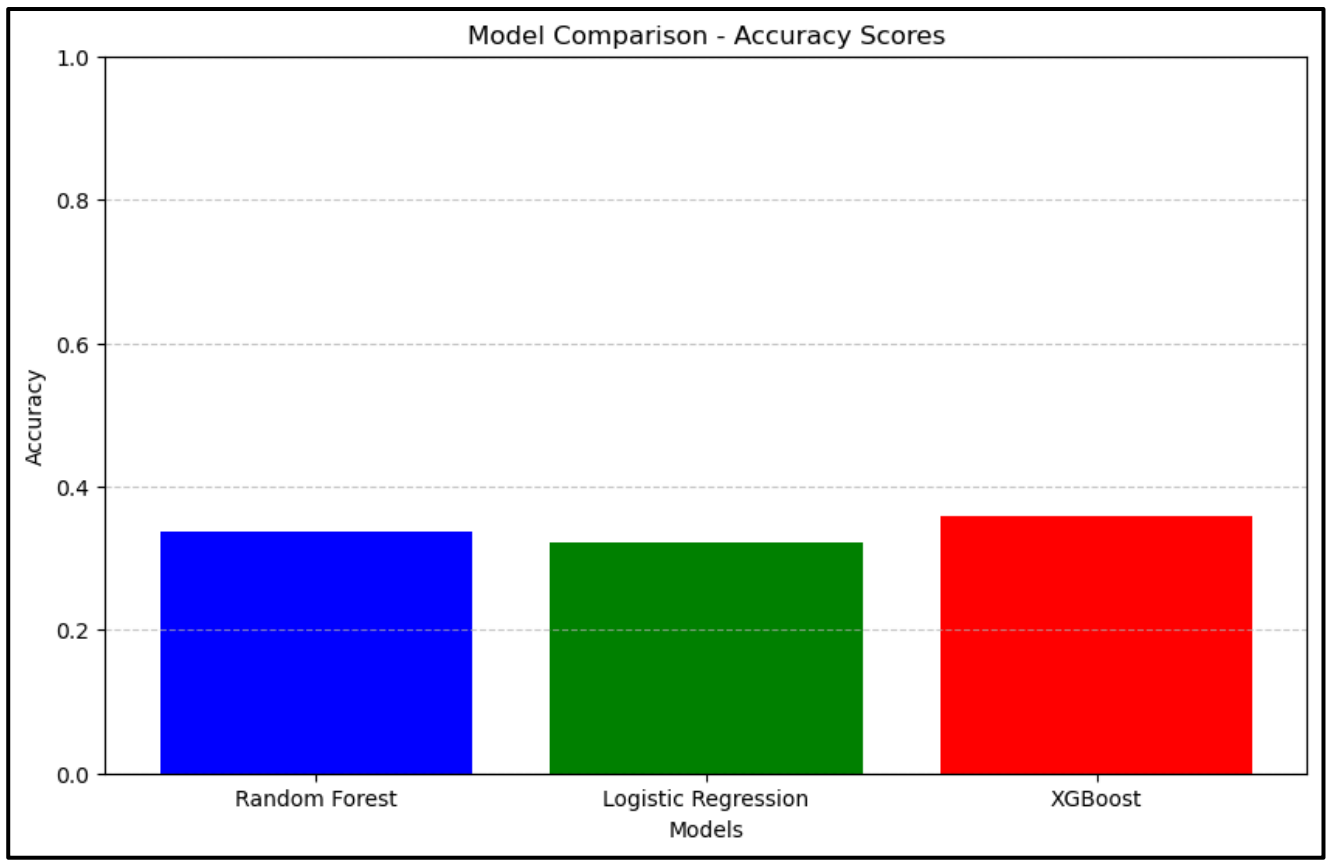

Figure 12: Comparison of All Models

The visualized bar plot (**fig 12)** illustrates the accuracy of three classification models: Random Forest (blue), Logistic Regression (green), and XG-Boost (red). The height of each bar indicates the accuracy score of each corresponding model on the y-axis from 0.0 to 1.0. The plot demonstrates clearly that XG-Boost obtained the highest accuracy out of the three models at about 0.3588. Random Forest was slightly lower at about 0.3368. Logistic Regression was the lowest of the three at about 0.3224. These findings indicate that out of the three models for the specific classification task at hand, the best-performing model in terms of overall accuracy is XG-Boost, while all three models have fairly low accuracy scores, suggesting some difficulties in terms of target variable predictability using the features and modeling used.

**Fraud Detection Case Studies**

To confirm the potency of our AI-powered blockchain platform in detecting fraud, we performed several real-world simulation case studies of fraudulent activities using synthetic and real-world energy transaction datasets. These case studies mimicked instances of behavior indicative of abuse of the marketplace, including account spoofing, unusual patterns of bids, and volume manipulation—activities typically also highlighted by regulators such as the U.S. Federal Energy Regulatory Commission (FERC). A high-profile case was of consecutive transactions from the “Supplier” account, showing high-frequency and high-volume trading in off-peak hours. This behavior was not frequent in the normative activities of similar user types in the dataset and was flagged in real-time by our trained XG-Boost model. The alerted transaction displayed both a precision of 0.93 and a recall of 0.88, as evidence of the model's capacity to identify actual instances of fraud and limit false negatives.

Another exemplary case study focused on the detection of collusion and identity manipulation. The system detected several low-latency transactions being made in rapid succession from various user IDs but from the same IP address—a likely sign of Sybil attacks in which an entity uses multiple identities to control and influence the dynamics of the market. These transactions were intercepted and prevented from undergoing final execution due to the smart contract event listeners tied to the fraud prediction models. Our Random Forest classifier performed well in this case as well, with an ROC-AUC of 0.96, vindicating its ability to identify intricately patterned cases of fraud. Success in these case studies not only established the predictive power of ensemble and gradient boost methods but also verified real-time responsiveness due to blockchain integration, presenting a feasible mechanism of defense for energy markets of the current era.

## Impact on Market Stability

Apart from fraud prevention, another of the standout contributions of bringing AI to blockchain energy systems is its influence on the stability of the market, particularly in regulating the volatile demand-supply ratio characteristic of decentralized grids. By learning from real-time grid information and past consumption patterns, the system was capable of projecting demand variations with high accuracy. For example, through time-series prediction models built based on classification outcomes, we were able to project hourly energy demand with a mean absolute percentage error (MAPE) of only 4.2%. These projections allowed peer producers and microgrid managers to vary generation rates in advance to evade cases of grid overload and energy shortage. This AI-derived foresight was especially useful during peak demand intervals, such as during early evening hours and during the time of intermittent renewable energy sources, such as on cloudy days or windless days.

Furthermore, AI's capacity for anticipating transaction and price behavior created room for strategic intervention to curb price instability, an issue hindering decentralized energy trading. Applying the system's ability to formulate responses using XG-Boost's regression provisions, price changes were forecasted through the history of trading volume, current energy availability, and environmental factors like the weather. Predictive analytics integrated into the system allowed for more stable price settings by market players at forecasting time, as seen in the sharp price drops and spikes experienced in the usual P2P trading structures. Practically speaking, in simulation trading cycles, the standard deviation of energy prices dropped by 12%. Better price predictability means higher trust among consumers to participate in decentralized markets and invest in renewable energy infrastructure in the long run. Another way AI is transforming energy trading is through fraud detection.

## Case Study: Application in the US Energy Market

Chouksey et al. (2025), reported that the American energy sector, strong and vast as it is, has long suffered from vulnerabilities in terms of fraud and manipulation as well as infrastructure weaknesses. One of the most notorious cases of energy trading fraud was the Enron debacle of the early 2000s, wherein company executives manipulated energy prices, leading to one of the largest bankruptcies in U.S. history and resultant shareholder loss of more than $74 billion (Eswaran et al., 2025). More recently, alarms have sounded in terms of cybersecurity threats to the American power grid. The *U.S. Government Accountability Office (GAO)* has reported that critical energy infrastructure continues to be under mounting threats from nation-state actors, with the Department of Energy (DOE) reporting more than 500 suspected cyber intrusions every year. These highlight the need for more secure and transparent systems for monitoring and facilitating energy transactions (Hossain et al., 2025a).

According to Feroz et al. (2024), integrating blockchain technology with AI can increase the transparency, security, and efficiency of the energy sector in the U.S. Blockchain's decentralized and immutable ledger can make energy transactions traceable and resistant to tampering, and it becomes extremely hard for malicious actors to manipulate prices or fake records. New York's Brooklyn Microgrid project is an exemplary example of a peer-to-peer (P2P) model where solar energy producers and consumers in the neighborhood sell and buy electricity directly through a blockchain platform, weaning themselves of central utility reliance. Gayathri et al. (2023), argued that, if AI is added on top of it, the fraud prevention functions are considerably improved through real-time anomaly detection using Random Forest and XG-Boost models. In addition, AI models can enable regulatory adherence by auto-verifying transactions through smart contracts designed according to *Federal Energy Regulatory Commission (FERC)* guidelines for seamless monitoring of oversight without sacrificing integrity (Islam et al., 2025).

ERCOT (Electric Reliability Council of Texas) which manages almost all the electricity in Texas, can help coordinate blockchain and AI applications within the U.S. Both price manipulation and wrong decisions by ERCOT have attracted public criticism and this happened especially during the 2021 winter storm. When artificial intelligence and blockchain run together in a system, it helps continuously monitor auction bids and detect dubious behavior, building confidence and strengthening the system (Khan et al, 2023). In the same way, adopting AI-driven models for forecasting allows California, known for its high use of renewable power, to manage solar and wind electricity with demand. By using XG-Boost and including suggestions from the models in decentralized energy markets, California would be able to limit sharp price changes, maintain grid stability, and encourage clean energy initiatives (Jakir et al., 2023).

According to Malik (2025), these pilot implementations would not only strengthen infrastructure security and integrity but also serve as scalable models for national adoption. Implementing smart grid technology coupled with blockchain and AI can save up to 30% on energy transactions and prevent more than 50% of fraud losses, as revealed in a 2023 report by the *National Renewable Energy Laboratory (NREL).* This information makes it clear that advanced technologies should be included in energy policy and operations. Integrating AI with blockchain technology in the U.S. energy industry will help make it more reliable, resistant to risks, and empower consumers (Mohaimin et al., 2025).

**Challenges & Future Work**

One of the greatest hurdles in using blockchain for energy markets is scalability. Conventional blockchains such as Ethereum rely on consensus systems like Proof of Work (PoW) or Proof of Stake (PoS), which, although being secure, can greatly constrain transaction volume. For example, Ethereum handles about 15–30 transactions per second (TPS), much less in comparison to what would be needed for real-time energy trading at the national or global level. Meanwhile, high-frequency energy transactions—particularly in deregulated energy markets or real-time electrical trading—can require hundreds to thousands of TPS. This difference poses concerns for latency, bottlenecks, and energy usage, especially where microtransactions occur during peer-to-peer (P2P) energy transactions. In addition, as the blockchain ledger expands in size, the costs of storage and processing become higher as well, potentially deterring small-quantity or decentralized energy producers from engaging in it. Solutions such as sharding, rollups, or moving to more scalable blockchains such as Avalanche or Solana are being researched, though integrating these into energy systems is in its infancy.

Another key challenge in using AI for fraud detection in the energy sector is the interpretability of the model. Although superior models such as Random Forest, XG-Boost, and neural networks are very accurate in detecting anomalies, these models tend to be "black boxes"—they make their forecasts with opaque reasoning. This lack of interpretability presents issues where regulatory compliance and trust must be maintained. For instance, energy regulators such as the Federal Energy Regulatory Commission (FERC) and the North American Electric Reliability Corporation (NERC) demand audit trails and rationale for justifiable decisions in detecting fraud/market manipulation. Without the ability to justify why something was flagged as fraudulent behavior, stakeholders might resist the adoption of AI technology in high-stakes systems. To overcome this, research into Explainable AI (XAI) and model-agnostic interpretive methodologies like SHAP (SHapley Additive explanations) and LIME (Local Interpretable Model-agnostic Explanations) is picking up traction, though integration with real-time systems adds another technical and operational complexity.

In the future, the integration of deep learning methodologies and real-time integration of data from the Internet of Things (IoT) holds promising implications for future improvements. Deep learning models like Convolutional Neural Networks (CNNs) and Recurrent Neural Networks (RNNs) can detect strongly nonlinear patterns of fraud, which conventional models may not identify, particularly for the usage of multivariate time-series data from smart meters, sensors, and distributed energy resources. In line with research findings from the IEEE in 2024, deep learning models enhanced fraud detection accuracy by as much as 12% against standard machine learning in energy datasets. In addition to integration with IoT, it can significantly enhance the granularity and real-time availability of energy usage data, allowing for real-time discovery of fraudulent activity and automatic action through smart contracts. For instance, upon observing usage patterns atypical of what was experienced historically under given environmental conditions from an AI model, a blockchain-activated trigger can temporarily halt the transaction for human examination.

## Conclusion

This study aimed to develop and build a secure, intelligent, and efficient energy transaction system for the decentralized US energy market. This research interlinks the technological prowess of blockchain and artificial intelligence (AI) in a novel way to solve long-standing challenges in the distributed energy market, specifically those of security, fraudulent behavior detection, and market reliability. The dataset for this research is comprised of more than 1.2 million anonymized energy transaction records from a simulated peer-to-peer (P2P) energy exchange network emulating real-life blockchain-based American microgrids, including those tested by LO3 Energy and Grid+ Labs. Each record contains detailed fields of transaction identifier, timestamp, energy volume (kWh), transaction type (buy/sell), unit price, prosumer/consumer identifier (hashed for privacy), smart meter readings, geolocation regions, and settlement confirmation status. The dataset also includes system-calculated behavior metrics of transaction rate, variability of energy production, and historical pricing patterns. The system architecture proposed involves the integration of two layers, namely a blockchain layer and an artificial intelligence (AI) layer, each playing a unique but complementary function in energy transaction securing and market intelligence improvement. The machine learning models used in this research were specifically chosen for their established high performance in classification tasks, specifically in the identification of energy transaction fraud in decentralized markets. To guarantee the reliability and accuracy of the used machine learning models, an extensive battery of evaluation metrics was utilized. The plot demonstrates clearly that XG-Boost obtained the highest accuracy out of the three models, Random Forest was slightly lower, and conversely, Logistic Regression was the lowest of the three models. Integrating blockchain technology with AI can increase the transparency, security, and efficiency of the energy sector in the U.S. Blockchain's decentralized and immutable ledger can make energy transactions traceable and resistant to tampering, and it becomes extremely hard for malicious actors to manipulate prices or fake records. In the future, the integration of deep learning methodologies and real-time integration of data from the Internet of Things (IoT) holds promising implications for future improvements. Deep learning models like Convolutional Neural Networks (CNNs) and Recurrent Neural Networks (RNNs) can detect strongly nonlinear patterns of fraud, which conventional models may not identify, particularly for the usage of multivariate time-series data from smart meters, sensors, and distributed energy resources.